\newcommand{\be}{\begin{equation}}
\newcommand{\ee}{\end{equation}}
\newcommand{\brr}{\begin{eqnarray}}
\newcommand{\err}{\end{eqnarray}}

\newcommand{\bd}{\begin{displaymath}}
\newcommand{\ed}{\end{displaymath}}

\newcommand{\bfig}{\begin{figure}}
\newcommand{\efig}{\end{figure}}

\documentclass{iopart}
\usepackage{ifvtex}
\usepackage{ifpdf}
\usepackage[dvips]{graphicx}
\usepackage{amssymb}
\usepackage{upmath}
\usepackage{mathrsfs}
\usepackage{upgreek}
\usepackage{yfonts}
\usepackage{dsfont}
\usepackage{relsize}
\usepackage{stackrel}

\usepackage{iopams}
\usepackage{graphicx}
\usepackage[all]{xy}
\begin{document}
\title{A squeezing formalism for finite dimensional quantum systems}
\author{M. Ruzzi}
\address{Navier Desenvolvimento e Tecnologia, \\ Av. Sete de Setembro, 4476, 80250-210, Curitiba, PR, Brazil}
\ead{maurizio@navier.com.br}
\begin{abstract}
This article presents a squeezing transformation for quantum systems associated to
finite vector spaces. The physical idea of squeezing here is taken from the action of the usual squeezing operator
over wave functions defined on a real line, that is, a transformation capable to diminish (or enhance) the mean square deviation of a
centered distribution. As it is discussed, the definition of such an operator on finite dimensional vector spaces is not a trivial matter, but, 
on the other hand, has obvious connections with problems such as spin squeezing and (finite) quantum state reproduction.
\end{abstract}
\pacs{03.65.-w,03.65.Ca,42.50.-p,42.50.Dv}

\section{\protect\bigskip Introduction}

\bigskip\ Quantum mechanics has a lot of peculiar features. One of them is
that it is able to cope with physical quantities which manifest themselves
in two quite different ways, namely those which are better described by
continuous variables (a case which we hereafter shall refer to as the
`continuum'), and those described by finite and discrete variables
(mentioned as the `discrete'). These both `versions' of quantum mechanics
have their own idiosyncratic features, and, despite the late rise of
interest in finite dimensional systems triggered mostly by the field of
quantum information (which in fact deals with both continuous and discrete
cases), one can safely say that the continuous case have a much richer (and
better established) variety of theoretical tools. Among the various examples
one could name here, a particularly sound one is that of squeezing. The
importance of the squeezing transformation (and the squeezed states) in
general quantum mechanics, quantum optics and, obviously, in quantum
information theory associated to continuous variables goes literally without
saying, and references on the subject are countless (some good examples,
among many, of textbooks and review articles are \cite{squeeztb} \cite
{squeezra}). On the other hand, however, even the idea of squeezing is
hardly touched upon in regard of finite dimensional quantum systems.

In the continuum, (that is, in the quantum mechanical description of the
usual, continuously labeled, canonically conjugated variables of position
and linear momentum) the unitary transformation of squeezing acts reducing
the dispersion related to, for example, the position observable (for
zero-mean distributions). The Fourier transform connection between the
wave-function in the position and momentum representations prevents from any
conflict with the conceptual content of the indeterminacy principle, as the
same transformation `stretches' the wave function in the dual momentum
representation. The domain of such wave-functions is dense and infinite, and
this both attributes are crucial in order to be possible to squeeze any
given wave-function. In rough terms, if the domain were not infinite, from
where the points (in the new, squeezed, representation) near the boundaries
of a given region (that can always be made arbitrarily large) would have
come from? And if the domain were not dense, to where all this points would
go? In other words, how can you squeeze a given wave function, if this wave
function is defined over a discrete and finite grid,\textit{\ keeping the
grid fixed}? All those \ ``ifs'' are obviously unnecessary concerns as long
as one is dealing only with the usual Cartesian coordinates, but lie in the
core of the problem if one would like to consider the problem of squeezing
in discrete finite dimensional quantum systems.

A well known squeezing operator in finite dimensional quantum systems would
have immediate consequences in the quantum information theory for such
systems. The \textit{\ no-cloning theorem} \cite{noc} asserts that quantum
states may only be completely reproduced if and only if they are an element
of a given known vector basis. Naturally, the higher the overlap of a given
state with some known basis element, the better such state can be cloned 
\cite{qsr}. Now, as we will discuss, a squeezing operator can be regarded as
an operator which reduces (or increases) the dispersion (the mean square
deviation from the mean) of any given state with respect to a given
observable. In other words, such operator increases (reduces) the overlap
between a given state and an untransformed element of a basis set, thus
enhancing (dimishing) the fidelity of the reproduction of such state.
Naturally, the actual implementation of any unitary transformation in a real
device have its own difficulties and limitations, which always have to be
taken into account. However, it is always better to face a technical
difficulty (the greater it may be) than a theoretical impossibility.

The aim of this article is thus to develop a mathematically consistent
approach for squeezing in finite dimensional quantum systems, introducing
what might be called squeezed discrete oscillator states and, naturally, the
squeezing operator. The unitarity of the squeezing operator here proposed is
a delicate issue. To put it in a paragraph, first a provisional squeezing
operator is presented, which is a genuine squeezing operator in the sense
that when properly applied it may reduce (or enhance) the mean square
deviation of any finite distribution. Such operator, however, is not
unitary. Unitarity is reached only in approximate fashion, so an objective
criterion of approximate unitarity is introduced in an appendix. Therefore,
unless explicitly stated otherwise, all use of the term ``unitary''
throughout this paper is in the approximate sense there defined. It must be
stressed that the approximations presented are extremely good.

In the forthcoming section II the difficulties above presented regarding
squeezing for finite dimensional quantum systems are discussed, and
necessary initial discussions in order to achieve a finite dimensional
squeezing operator are made, including a brief outlook of known properties
of squeezing in continuous systems, which is absolutely necessary to guide
the discussion. Section III is devoted to the (wayward) definition of the
squeezing operator. The actual squeezing is presented in section IV, while
final remarks and discussions are left to section V.

\section{The theoretical problem of squeezing in finite dimensional quantum
systems}

In quantum mechanics of usual, infinite dimensional and continuously labeled
Hilbert spaces, the squeezing operator acts over the position eigenstates as 
\begin{equation}
\mathbf{S}(\lambda )|x\rangle =\frac{1}{\sqrt{\lambda }}|\lambda x\rangle ,
\end{equation}
where $\lambda $ is a real number which parametrizes the transformation. It
is not hard to see the scaling effect that this transformation has on a
given wave function, as a state represented by $\psi (x)$ would be
transformed in to one represented by $\psi (\lambda ^{-1}x)$, thus the name
squeezing (stretching, if $\lambda $ is larger than one). However, in a
finite, discrete domain, an operator which maps a state $|k\rangle $ onto $%
|\lambda k\rangle $ (with $k$ some discrete label of a given finite family
of eigenstates) hardly could be obtained or even interpreted as a squeezing
operator. In such domain, if $\lambda $ is larger than one, $\lambda k$
would undoubtedly fall off the domain of labels at least for some values of $%
k$, and if $\lambda $ is smaller than one, as the domain is not dense, $%
\lambda k$ would not be in the domain for some values (even all) of $k$
either. A more sophisticated possibility would be to consider the product \ $%
\lambda k$ as a modulo $N$ operation, as considered in \cite{vourdas}. In
this case the difficulties just pointed out surely do not arise (at least
for\ integer $\lambda $\ and prime $N$). However, such a transformation
cannot be interpreted as squeezing. It is a well known fact from number
theory that, given a set of integers $D$ obeying $\left\{ k\in D|0\leq k\leq
N-1\right\} $ (which is a complete set of residues modulo $N),$ the set $%
\left\{ \lambda k \, (\mbox{mod} N)\right\} ,$ for integer $\lambda $, coincides
with $D$, but in some different order \cite{numb}. In other words, this is
rather a shuffling transformation than a squeezing (or stretching) one. The
shape of an arbitrary wave function would be complete and radically altered
by such transformation, and, more importantly, the effect of this
transformation on the dispersion would depend heavily on the initial wave
function.

Thus the action of the squeezing operator over the families of eigenstates
of position and momentum gives no valuable hint of how an equivalent
operator in a discrete domain would act, and we must look elsewhere if we
want to find a good starting point for the discrete version of squeezing.

\subsection{Oscillator states, width and Fourier transforms in the continuous%
}

The eigenstates of position and momentum are not the only way to look at the
problem of squeezing. In a Schroedinger inspired approach to the issue, in
this sub-section we look at some simple (and well, well known) properties of
the harmonic oscillator states. This seemingly useless \ (but brief)
repetition of general knowledge will prove itself more than pertinent in the
next sections, as all expressions there obtained will end up being discrete
analogs of the ones here presented.

The ground state of the harmonic oscillator is described by a Gaussian
function of unit width (in natural units), 
\begin{equation*}
|0\rangle =\frac{1}{\mathcal{N}^{0}}\int_{-\infty }^{\infty }dx\exp \left[ -%
\frac{1}{2}x^{2}\right] |x\rangle ,\qquad \mathcal{N}^{0}=\pi ^{1/4},
\end{equation*}
written in terms of the position eigenstates $|x\rangle $. The whole set of
states is defined through the Hermite polynomials, 
\begin{equation*}
|n\rangle =\frac{1}{\mathcal{N}^{n}}\int_{-\infty }^{\infty }dxH_{n}(x)\exp %
\left[ -\frac{1}{2}x^{2}\right] |x\rangle ,\qquad \mathcal{N}^{n}=\sqrt{%
2^{n}n!}\pi ^{1/4}.
\end{equation*}
In order to account for different widths of the harmonic oscillator wave
functions, it is possible to define 
\begin{equation*}
|n;\lambda \rangle =\frac{1}{\mathcal{N}^{n}\sqrt{\lambda }}\int_{-\infty
}^{\infty }dxH_{n}\left( \frac{x}{\lambda }\right) \exp \left[ -\frac{1}{%
2\lambda ^{2}}x^{2}\right] |x\rangle .
\end{equation*}
In the momentum representation, the same states are represented by the
(integral) Fourier transforms of the functions above 
\begin{equation}
|n;\lambda \rangle =\frac{\sqrt{\lambda }}{\mathcal{N}^{n}}\int_{-\infty
}^{\infty }dpH_{n}\left( p\lambda \right) \exp \left[ -\frac{\lambda ^{2}}{2}%
p^{2}\right] |p\rangle ,  \label{repP}
\end{equation}
which have their widths values inverted in the process.

In an algebraic sense, if the \emph{Fourier operator} maps a position
eigenstate on a momentum eigenstate as 
\begin{equation*}
\mathcal{F}|x\rangle =|p\rangle ,
\end{equation*}
it directly follows that 
\begin{equation}
\mathcal{F}|n;\lambda \rangle =i^{n}|n;1/\lambda \rangle ,  \label{foun}
\end{equation}
and the oscillator states with unit width are eigenstates of $\mathcal{F}$.
By its turn, the squeezing operator acts over the unit width ground state as 
\begin{equation*}
\mathbf{S}(\lambda )|0;1\rangle =\frac{1}{\mathcal{N}^{0}\sqrt{\lambda }}%
\int_{-\infty }^{\infty }dx\exp \left[ -\frac{1}{2\lambda ^{2}}x^{2}\right]
|x\rangle =|0;\lambda \rangle ,
\end{equation*}
altering the width for the value $\lambda .$ In fact, in general 
\begin{equation*}
\mathbf{S}(\lambda )|n\rangle =|n;\lambda \rangle .
\end{equation*}
and, as the parameter $\lambda $ controls the variance associated to all
oscillator states, the state is therefore ``squeezed'' (for $\lambda <1$).
One must not forget that such ``squeezing'' may actually be a stretching if
one is looking at the momentum representation, as shown by Eq.(\ref{repP}). $%
\ $

Finally, oscillator states are complete and orthogonal, 
\begin{equation}
\langle n^{\prime };\lambda |n;\lambda \rangle =\delta _{n,n^{\prime
}}\qquad \sum_{n=0}^{\infty }|n;\lambda \rangle \langle n;\lambda |=\mathbf{%
\hat{1},}
\end{equation}
what allows us to write the squeezing operator as 
\begin{eqnarray}
\mathbf{S}(\lambda ) &=&\mathbf{S}(\lambda )\sum_{n=0}^{\infty }|n;1\rangle
\langle n;1| \\
\mathbf{S}(\lambda ) &=&\sum_{n=0}^{\infty }|n;\lambda \rangle \langle n;1|.
\label{sqz_number}
\end{eqnarray}

\subsection{Discrete oscillator states}

\subsubsection{Preliminaries}

Let the set $\left\{ |u_{k}\rangle \right\} _{k=-\ell }^{\ell }$ be some
complete and physically meaningful set of states\cite{schw,ruzzi1}, spanning
a $N$ dimensional vector space $\frak{V}$, associated with a given physical
quantity $O,$ so that \ 
\begin{equation*}
\mathbf{O}|u_{k}\rangle =o(k)|u_{k}\rangle ,
\end{equation*}
where $\mathbf{O}$ is the\ Hermitian operator representing the physical
quantity, with its associated set $\left\{ o(k)\right\} _{k=-\ell }^{\ell }$
of \ eigenvalues standing for the possible outcomes of physical measures of $%
O$. Defining a ``coordinate'' operator for such discrete set of states as $%
\mathbf{Q}|u_{k}\rangle =k|u_{k}\rangle ,$ the dispersion $\sigma $\
associated to any state may be written as 
\begin{equation*}
\sigma =\langle \mathbf{Q}^{2}\rangle -\langle \mathbf{Q}\rangle ^{2}.
\end{equation*}
If a given physical system is in a state $|u_{j_{0}}\rangle $ associated to
the value $o(j_{0})$ of $O$, the average value of the ``coordinate
operator'' \ $\mathbf{Q}$ is obviously $j_{0}$ and the dispersion is zero,
that is, the distribution is sharply peaked at $j_{0}$.

Let us also introduce, with the aid of the Discrete Fourier Transform (DFT),
the complementary set 
\begin{equation*}
|v_{k}\rangle =\sum_{j=-\ell }^{\ell }\exp \left[ -\frac{2\pi i}{N}kj\right]
|u_{j}\rangle ,
\end{equation*}
such that a given state $|\psi \rangle $, represented by a $\psi (k)$ wave
function in the $\{|u_{k}\rangle \}$ representation, is represented by the
DFT $\bar{\psi}(k)$ in the $\{|v_{k}\rangle \}$ representation. It is not
hard to see that these two representations play in the discrete the role
played by $\{|x\rangle \}$ and $\{|p\rangle \}$ in the continuum \cite
{ruzzi1}.

\subsubsection{Width parametrized eigenfunctions of the DFT}

Now we leap in to much lesser known grounds. Reference \cite{eujmp} presents
some very eloquent results regarding the discrete Fourier transform (DFT) of
Jacobi Theta functions, that completely parallel the ones presented in
subsection II.a. In particular, there it is shown that the set of functions 
\begin{equation}
f_{n}(j,\xi )=\sqrt{\frac{N}{\xi }}\left. \frac{\partial ^{n}}{\partial t^{n}%
}\vartheta _{3}\left( \frac{j}{N}-\frac{\epsilon }{\pi }\xi t,\frac{i\xi ^{2}%
}{N}\right) \exp \left[ t^{2}\right] \right| _{t=0},  \label{forma_diog}
\end{equation}
behaves, under DFT, as 
\begin{equation}
\frac{1}{\sqrt{N}}\sum_{j=0}^{N-1}\exp \left[ -\frac{2\pi i}{N}jk\right]
f_{n}(j,\xi )=i^{n}f_{n}(k,\xi ^{-1}).  \label{mg}
\end{equation}
The Jacobi $\vartheta _{3}$-function explicit form is \cite{vilenkin} 
\begin{equation*}
\vartheta _{3}\left( z,\tau \right) =\sum_{\alpha =-\infty }^{\infty }\exp 
\left[ i\pi \tau \alpha ^{2}\right] \exp \left[ 2\pi i\alpha z\right] ,
\end{equation*}
and with the help of the Hermite polynomials generating function, the $%
f_{n}(j,\xi )$ functions may also be put in the equivalent forms 
\begin{eqnarray*}
f_{n}(j;\xi ) &=&\frac{1}{N}\sum_{\alpha =-\infty }^{\infty }\exp \left[ -%
\frac{\pi }{N\xi ^{2}}(\alpha N+j)^{2}\right] H_{n}\left( \frac{\epsilon }{%
\xi }(\alpha N+j)\right) ,\qquad \epsilon =\sqrt{\frac{2\pi }{N}} \\
f_{n}(j;\xi ) &=&\frac{\epsilon }{N\sqrt{\xi }}(-i)^{n}\sum_{\alpha =-\infty
}^{\infty }\exp \left[ -\frac{\pi }{N}\xi ^{2}\alpha ^{2}+\frac{2\pi i}{N}%
j\alpha \right] H_{n}\left( \epsilon \xi \alpha \right) .
\end{eqnarray*}
These functions are periodic, that is, $f_{n}(j+N;\xi )=f_{n}(j;\xi )$. As
the second argument of the Theta functions controls their width \cite{eujmp}%
, expression (\ref{forma_diog}) ensures that the parameter $\xi $ controls
the dispersion associated to those functions, as it is also evident from
figures 1 and 2. The DFT preserves the analytic form of the functions $%
f_{n}(j;\xi )$, mapping $\xi $ into $\xi ^{-1}$, that is, a given width
value into its inverse, in close analogy to the behavior of the Gaussians
under integral Fourier transforms. The $f_{n}(j;\xi )$ functions are
symmetric about zero, and thus have zero mean in the sense that always 
\begin{equation}
\sum_{j=-\ell }^{\ell }j\ \left| f_{n}(j,\xi )\right| ^{2}=0,\qquad \ell =%
\mathsmaller{\frac{N-1}{2}}.  \label{zeromean}
\end{equation}

Therefore, one may define \emph{discrete oscillator states} as 
\begin{equation}
|n;\xi \rangle =\frac{1}{\mathcal{N}_{n,\xi }}\sum_{k=-\ell }^{\ell
}f_{n}(k;\xi )|u_{k}\rangle ,\qquad n\in \lbrack 0,N-1],  \label{dos}
\end{equation}
where $\mathcal{N}_{n,\xi }$ is the appropriate normalization factor. From
equation (\ref{zeromean}) it follows that $\langle n;\xi |\mathbf{Q|}n;\xi
\rangle =0.$\ By stating that ``the parameter $\xi $ controls the
dispersion'' we mean that $\sigma ,$ when calculated with respect to a
discrete oscillator state, is a monotonically increasing function of $\xi $. It is evident 
from the definitions that the ``coordinate'' operator $\mathbf{Q}$ is NOT diagonal in the discrete 
oscillator states representation. 

Although it is obvious, it must be kept clear that the only analogy used here is the one between the discrete oscillator states 
and the usual continuous harmonic oscillator eigenstates, and NOT general number states in a Fock space.  

Algebraically, if the Fourier operator in the discrete domain is $\mathbf{%
\mathcal{F}}_{d}|u_{k}\rangle =|v_{k}\rangle ,$ it is then immediate to
verify that 
\begin{equation}
\mathbf{\mathcal{F}}_{d}|n;\xi \rangle =|n;\xi ^{-1}\rangle ,
\label{invfoud}
\end{equation}
in analogy to Eq.(\ref{foun}). And, equivalently, 
\begin{equation*}
|n;\xi \rangle =\frac{1}{\mathcal{N}_{n,\xi ^{-1}}}\sum_{k=-\ell }^{\ell
}f_{n}(k;\xi ^{-1})|v_{k}\rangle ,\qquad n\in \lbrack 0,N-1],
\end{equation*}
that is, the state\ $|n;\xi \rangle $ is represented by $f_{n}(k;\xi ^{-1})$
in the dual $\{|v_{k}\rangle \}$ representation, again in a perfect analogy
with the continuous case.

\begin{figure}[t]
\hspace{1.5 em} 
\begin{minipage}[t]{\linewidth}
\includegraphics[angle=00,width=12cm]{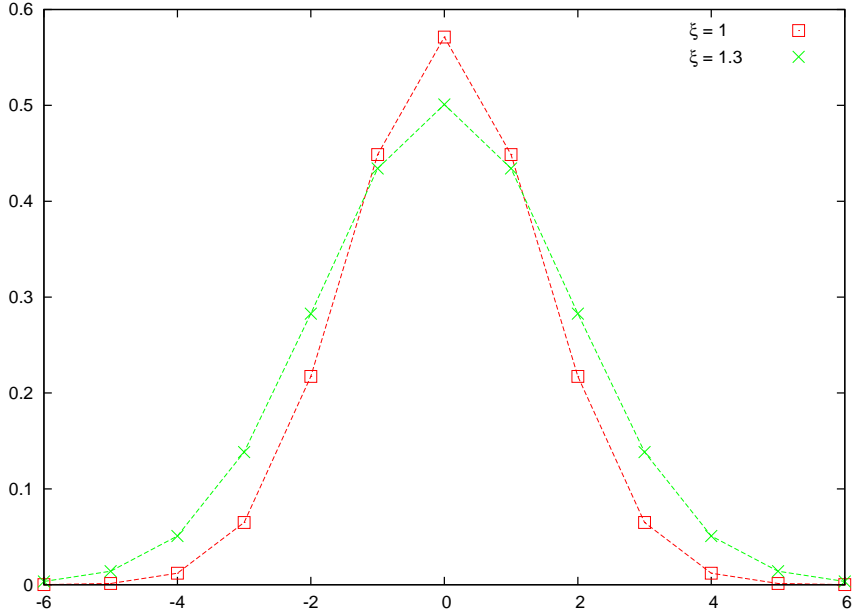}
\end{minipage} 
\caption{Plot of the functions $f_{0}(k;1)$ and $f_{0}(k;1.3)$. As those
functions are defined on integers, the dotted lines only guide the eye. Two
characteristics are explicit: First, the resemblance with the continuous
harmonic oscillator functions, and the increase in width when one goes from
the squeezing parameter $1$ to $1.3$. The parameters are choosen so that the squeezing effect is 
visible.}
\end{figure}

It might be safely assumed that, for odd $N$, the first $N$ discrete
oscillator states are complete\cite{mehta}. For even $N$, the $f_{n}(k;1)$
functions have some interesting (and still not fully comprehended)
properties, and the function associated to the upper bound $N-1$, $%
f_{N-1}(k;1),$ is identical to $f_{N-5}(k;1)$ up to a minus sign. In view of
that, for even dimensions we must substitute $f_{N-1}(k;1)$ by $f_{N+3}(k;1)$
(as $i^{N-1}=i^{N+3})$ in order to have a complete set. In the following,
for simplicity, we restrict our notation to the odd case.

A little digression on the unit width states $\{|n;1\rangle \}$ is
necessary. The first important question to address is if this set is
complete. First, in reference \cite{mehta}, Mehta conjectures that this set
of states is linearly independent (and therefore complete). Such conjecture
is heavily supported by numerical evidence \footnote{
In reference \cite{eujmp} there is a misjudgment of Mehta's conjecture.
Guided by his own concerns on the subject, the author assumed that Mehta has
conjectured on the orthogonality of the $f_{n}(k;1)$ functions, while the
actual conjecture is about the linear independence of the set.}. We present
the actual values of the overlap matrix $\langle n^{\prime };1|n;1\rangle ,$
for the case $N=13$, which are well illustrative.\ As those states are
eigenstates of the Fourier operator with eigenvalue $i^{n},$ $\langle
n;1|n^{\prime };1\rangle $ is exactly zero if $i^{n}\neq i^{n^{\prime }}$.
Non-zero off-diagonal matrix elements occur for $|n-n^{\prime }|$ a multiple
of $4$. In fact, in general, only for the larger values of the indices $n$
and $n^{\prime }$ the values of the off-diagonal elements become significant 
\cite{natig}. Reference \cite{eujmp} presents a detailed study of the inner
product $\langle n;1|n^{\prime };1\rangle $. \ If the width $\xi $ is
different than $1$, there are more non-zero off diagonal elements. However,
for values of $\xi $\ in the vicinity of $1$ $(0.9\lesssim \xi \lesssim 1.1)$
the deviation from the values obtained for $\xi =1$ is small. In fact, for
the lower values of the index $n$, the inner product $\langle n;\xi
|n^{\prime };\xi \rangle $ does not (significantly) depend on $\xi $, and
therefore a transformation mapping $|n^{\prime };1\rangle $ on to $%
|n^{\prime };\xi \rangle $ is (approximate to) an unitary one, a fact which
is duly explored in \cite{maruga2}. A precise notion of
``approximate unitarity'' is presented in the appendix.

\begin{figure}[t]
\hspace{2.5 em} 
\begin{minipage}[t]{\linewidth}
\includegraphics[angle=0,width=11cm]{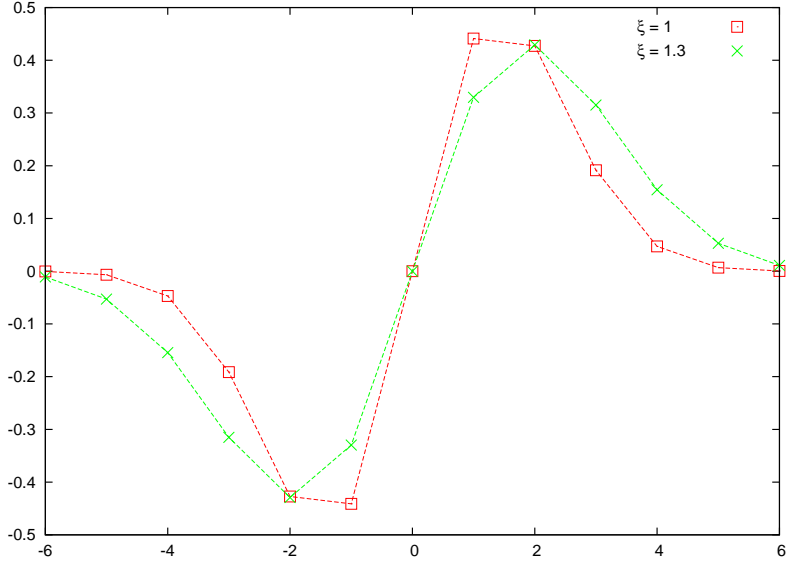}
\end{minipage} 
\caption{Plot of the functions $f_{1}(k;1)$ and $f_{1}(k;1.3)$. The dotted
lines only guide the eye.}
\end{figure}

\section{\protect\bigskip The squeezing operator for finite dimensional
quantum systems}

\bigskip The results of the previous section show the deep analogy between
discrete and continuous oscillator states, as both families are parametrized
by its width value, which is inverted under Fourier transformation, not to
mention the shape of the functions itself, as can be seen in figures 1 and
2. In fact, one may obtain the discrete $f_{n}(k;\xi )$ from the continuous
oscillator functions through the Poisson summation technique \cite{bellman}.
Such results give us a path to follow in order to define a squeezing
operator suitable for finite dimensional quantum systems.

\subsection{Preliminary attempt}

The general idea is that the eigenstates of the harmonic oscillator provide
a ``natural'' representation of the squeezing operator, and that discrete
oscillator states provide a discrete analog of that representation on finite
dimensional quantum systems. So, bearing in mind expression \ref{sqz_number}%
, we can define an finite dimensional squeezing operator of the form 
\begin{equation}
\mathbf{\Xi }_{p}(\xi )=\sum_{n=0}^{N-1}|n;\xi \rangle \langle n;1|.
\label{S}
\end{equation}
where the subscript $p$ stands for 'provisional'. Now, let us suppose for a
moment that the set $\left\{ |n;\xi \rangle \right\} $ is orthogonal for all
values $\xi $, or at least for a reasonable range around $1.$ Such an
operator, in that case, would perfectly parallel the continuous squeezing
operator. By construction, it would squeez (stretch) a $|m;1\rangle $ state
characterized by width $1$ to a smaller (higher) value $\xi $, that is 
\begin{equation*}
\mathbf{\Xi }_{p}(\xi )|n;1\rangle =|n;\xi \rangle ,
\end{equation*}
just like in the continuous case. Also, it would follow (after some algebra)
that $\mathbf{\Xi }_{p}^{\dagger }(\xi )=\mathbf{\Xi }_{p}(\xi ^{-1}),$
what, if $\mathbf{\Xi }_{p}$ is unitary, leads to the convenient property 
\begin{equation*}
\mathbf{\Xi }_{p}(\xi ^{-1})=\mathbf{\Xi }_{p}^{-1}(\xi ).
\end{equation*}
The unitarity of the $\mathbf{\Xi }_{p}(\xi )$ operator would follow
directly 
\begin{eqnarray*}
\mathbf{\Xi }_{p}(\xi )\mathbf{\Xi }_{p}^{\dagger }(\xi )
&=&\sum_{n=0}^{N-1}\sum_{m=0}^{N-1}|n;\xi \rangle \stackrel[\delta _{m,n}]{}{%
\underbrace{\langle n;1|m;1\rangle }}\langle m;\xi | \\
\mathbf{\Xi }_{p}(\xi )\mathbf{\Xi }_{p}^{\dagger }(\xi )
&=&\sum_{n=0}^{N-1}|n;\xi \rangle \langle n;\xi |=\mathbf{\hat{1},}
\end{eqnarray*}
\begin{eqnarray*}
\mathbf{\Xi }_{p}^{\dagger }(\xi )\mathbf{\Xi }_{p}(\xi )
&=&\sum_{n=0}^{N-1}\sum_{m=0}^{N-1}|n;1\rangle \stackrel[\delta _{m,n}]{}{%
\underbrace{\langle n;\xi |m;\xi \rangle }}\langle m;1| \\
\mathbf{\Xi }_{p}^{\dagger }(\xi )\mathbf{\Xi }_{p}(\xi )
&=&\sum_{n=0}^{N-1}|n;1\rangle \langle n;1|=\mathbf{\hat{1},}
\end{eqnarray*}
where the completeness of the $\left\{ |n;\xi \rangle \right\} $ states is
explicitly used.

Under the orthogonality assumption, then, it is true that the above operator
is unitary, and squeezes, or stretches, the discrete oscillator states,
which do form a complete set in which any other state might be expanded. The
squeezing, or the stretching, is understood in the sense that the dispersion 
$\sigma $ is diminished, or enhanced. However, as stated above, the discrete
oscillator states are, in general, not orthogonal. Therefore, we must amend
our initial (and naive) approach.

\begin{table}[t]
\centering\par
\begin{center}
\hspace*{-3 em}
\thicklines
\vline
\begin{tabular}{ccccccccccccc}
1 & 0 & 0 & 0 & 0.00 & 0 & 0 & 0 & 0.00 & 0 & 0 & 0 & 0.00 \\ 
0 & 1 & 0 & 0 & 0 & 0.00 & 0 & 0 & 0 & 0.00 & 0 & 0 & 0 \\ 
0 & 0 & 1 & 0 & 0 & 0 & 0.00 & 0 & 0 & 0 & 0.00 & 0 & 0 \\ 
0 & 0 & 0 & 1 & 0 & 0 & 0 & 0.00 & 0 & 0 & 0 & 0.05 & 0 \\ 
0.00 & 0 & 0 & 0 & 1 & 0 & 0 & 0 & 0.00 & 0 & 0 & 0 & 0.07 \\ 
0 & 0.00 & 0 & 0 & 0 & 1 & 0 & 0 & 0 & 0.05 & 0 & 0 & 0 \\ 
0 & 0 & 0.00 & 0 & 0 & 0 & 1 & 0 & 0 & 0 & 0.01 & 0 & 0 \\ 
0 & 0 & 0 & 0.00 & 0 & 0 & 0 & 1 & 0 & 0 & 0 & -0.67 & 0 \\ 
0.00 & 0 & 0 & 0 & 0.00 & 0 & 0 & 0 & 1 & 0 & 0 & 0 & 0.42 \\ 
0 & 0.00 & 0 & 0 & 0 & 0.05 & 0 & 0 & 0 & 1 & 0 & 0 & 0 \\ 
0 & 0 & 0.00 & 0 & 0 & 0 & 0.01 & 0 & 0 & 0 & 1 & 0 & 0 \\ 
0 & 0 & 0 & 0.05 & 0 & 0 & 0 & -0.67 & 0 & 0 & 0 & 1 & 0 \\ 
0.00 & 0 & 0 & 0 & 0.07 & 0 & 0 & 0 & 0.42 & 0 & 0 & 0 & 1
\end{tabular}
\thicklines
\vline
\end{center}
\par
\hfill
\par
\begin{picture}(100,0)
\put(-72,20){\dashbox{0.5}(0,155){}}
\put(-158,141){\dashbox{0.5}(388,0){}}
\end{picture}
\caption{Overlap matrix for $N=13$ and $\protect\xi =1$. Values presented
with two decimal digits were numerically calculated. The blocks were chosen
so as the first diagonal block is the identity matrix, and the off-diagonal
blocks have only elements smaller than $0.01$}
\end{table}

\subsection{First attempt: Non-orthogonal decomposition}

First of all, as the set $\left\{ |n;\xi \rangle \right\} _{n=0}^{N-1}$ is
not an orthogonal set, in order to work properly with it we must adopt the
usual techniques valid for non-orthogonal basis sets \cite{arta}. Due to the
non-orthogonality, the operator defined on \ref{S} does \textit{not} map a
unit width discrete oscillator state in to another discrete oscillator state
of different width. This must be accomplished by means of a dual set $%
\left\{ |m;\xi )\right\} _{m=0}^{N-1}$ with the property 
\begin{equation*}
(m;\xi |n;\xi \rangle =\delta _{m,n}.
\end{equation*}
In this form the closure relation may assume one of the forms 
\begin{equation*}
\sum_{n=0}^{N-1}|n;\xi \rangle (n;\xi |=\sum_{n=0}^{N-1}|n;\xi )\langle
n;\xi |=\mathbf{\hat{1},}
\end{equation*}
so the decomposition of a given state $|\psi \rangle $ would read, for
example, 
\begin{equation*}
|\psi \rangle =\sum_{n=0}^{N-1}|n;\xi \rangle (n;\xi |\psi \rangle
=\sum_{n=0}^{N-1}\psi _{n}^{\xi }\ |n;\xi \rangle .
\end{equation*}
It is worth emphasizing that such a dual set always exists if the overlap
matrix defined by the elements $\langle j;\xi |l;\xi \rangle $ has an
inverse. Armed with the above results one could study the properties of the
given operator 
\begin{equation*}
\mathbf{\Xi }(\xi )=\sum_{n^{\prime}
=0}^{N-1}|n^{\prime}
;\xi \rangle (n^{\prime}
;1|.
\end{equation*}
Now, the action of this operator over the discrete oscillator states is
clearly $\mathbf{\Xi }(\xi )|n;1\rangle =|n;\xi \rangle .$ Therefore, in
fact it squeezes, or stretches, the discrete oscillator states. It is
interesting to note that the operator 
\begin{equation*}
\mathbf{\bar{\Xi}}(\xi )=\sum_{n^{\prime}=0}^{N-1}|n^{\prime}
;1 \rangle (n^{\prime};\xi|
\end{equation*}
satisfies the relation 
\begin{equation*}
\mathbf{\Xi }(\xi )\mathbf{\bar{\Xi}}(\xi )=\mathbf{\bar{\Xi}}(\xi )\mathbf{%
\Xi }(\xi )=\mathbf{\hat{1},}
\end{equation*}
thus $\mathbf{\bar{\Xi}}(\xi )=\mathbf{\Xi }^{-1}(\xi )\mathbf{.}$

The condition for unitarity would then be $\mathbf{\Xi }^{\dagger }(\xi )=%
\mathbf{\bar{\Xi}}(\xi )$, which in general does not hold.

\subsection{Second attempt: Squeezing conjecture and approximate unitarity}

The non-orthogonality of the $|n;\xi \rangle $ states is a serious hindrance
for the construction of an unitary squeezing operator, but it can be at
least approximately circumvented. From now on, we follow the criteria
defined on the appendix for \ ``approximate unitarity'', and all further
references to ``unitarity'' are made in this approximate sense.

Some basic definitions are necessary: For a given dimensionality $N,$ let $%
\mathcal{V}_{l}^{\xi }$ be the subspace spanned by the set of the first $%
N_{l}$ (where $N_{l}$ shall be defined \textit{a posteriori}) discrete
oscillator states, $\left\{ |m;\xi )\right\} _{m=0}^{m=N_{l}-1},$ and $%
\mathcal{V}_{h}^{\xi }$ the ``remaining'' subspace (orthogonal to $\mathcal{V%
}_{h}^{\xi }$) so that $\mathcal{V}_{l}^{\xi }\oplus \mathcal{V}_{h}^{\xi }=%
\mathcal{V.}$ Accordingly, $\mathbf{\hat{1}}_{l}$ and $\mathbf{\hat{1}}_{h}$
are the respective identity operators on those subspaces.

In that case we can then define an \emph{unitary} operator of the form 
\begin{equation}
\Xi _{u}(\xi )=\sum_{n=0}^{N_{l}-1}\ |n;\xi \rangle (n;1|+\mathbf{\hat{1}}%
_{h},
\end{equation}
where $\mathbf{\hat{1}}_{h}=\sum_{n=N_{l}}^{N-1}\ |n;\xi \rangle (n;\xi |$.
Now, such an operator still has a lot of features which parallel those of
the continuous squeezing operator. By construction, it squeezes (stretches)
a $|m;\xi \rangle $ state characterized by width $1$ to a smaller (higher)
value $\xi $, that is 
\begin{equation*}
\Xi _{u}(\xi )|n;1\rangle =|n;\xi \rangle ,
\end{equation*}
provided that $n<N_{l}$ (otherwise it leaves the state unchanged).\ It is
not hard to see that $\Xi _{u}(\xi )$ will be unitary as long as 
\begin{equation*}
(m;1|n;\xi \rangle \approx 0,\qquad \mbox{for }m<N_{l}\leq n.
\end{equation*}

In order to put forth this approximation, for a given dimensionality $N$ of
the original state space, one identifies, as in done in table I, two
diagonal blocks of the overlap matrix $\langle n;\xi |n^{\prime };\xi
\rangle ,$ where the lower block is the identity matrix. The blocks must be
chosen so that all the elements in the non-diagonal blocks are negligible.
The first block dimensions are $N_{l}\times N_{l}$, whereas the remaining
block is $N_{h}\times N_{h},$ obviously with $N_{l}+N_{h}=N$. The appendix
gives more details on this issue, but the general idea is the one described
above.

We are now in position to make a \textit{squeezing conjecture: }The unitary
operator which squeezes the discrete oscillator states belonging to $%
\mathcal{V}_{l}^{\xi }$ squeezes all states defined on that subspace. 
\textit{\ }

\section{Actual Squeezing}

Let us then briefly see the action of the finite dimensional squeezing
operator. As the set $\{|n;\xi \rangle \}_{n=0}^{N-1}$ is complete, an 
\textit{arbitrary} state can always be described by 
\begin{equation}
|\varphi \rangle =\sum_{n=0}^{N-1}\varphi (n;\xi )|n;\xi \rangle .
\label{decomp}
\end{equation}
One can, in principle, arbitrarily choose the value of $\xi $ in
decomposition (\ref{decomp}). Choosing then $\xi =1,$ the action of $\mathbf{%
\Xi }_{u}(\xi )$ over $|\varphi \rangle $ may be simply written as 
\begin{equation*}
\mathbf{\Xi }(\xi )|\varphi \rangle =\mathbf{\Xi }(\xi
)\sum_{n=0}^{N-1}\varphi (n;1)|n;1\rangle =\sum_{n=0}^{N_{l}-1}\varphi
(n;1)|n;\xi \rangle +\sum_{n=N_{l}}^{N-1}\varphi (n;1)|n;1\rangle .
\end{equation*}
In figure 3 we present the actual squeezing of a square wave function. A
number of different examples could have been shown, but their behaviour is
basically the same. Once the distribution is roughly centered around zero,
and can be reasonably described by the lower discrete oscillator states, the
effect of the squeezing transformation on it will be appreciable. In the
particular case here shown, the squeezing effect can be easily grasped from
the figure.

\begin{figure}[t]
\hspace{2 em} 
\begin{minipage}[t]{\linewidth}
\includegraphics[angle=00,width=12cm]{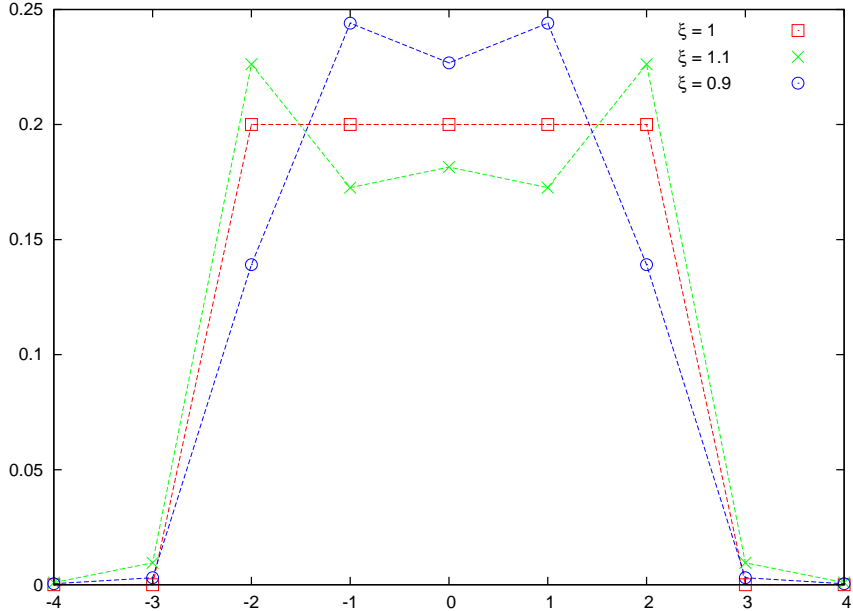}
\end{minipage} 
\caption{Square wave function before and after squeezing transformations of
parameters 1.1 and 0.9}
\end{figure}

In fact, although the squeezing formalism here presented is clearly built to
work for zero-mean distributions, one could easily generalize it simply by
adopting conveniently shifted discrete oscillator functions, or making use
of the well known unitary displacement operators for finite dimensional
systems \cite{schw,weyl}.

The effectiveness of the result shown here should not be underestimated due to the presentation of only one example. Any centered, bell-shaped distrution would be even easyer handed by the discrete squeezing operator, the square wave function was choosen exactly because it is very different from the discrete oscillator states (and the discrete squeezing operator, by definition, correctly squeezes those states). The squeezing operator
is able to unitarily control the dispersion of any centered distribution.

\section{Final remarks}

A straightforward analogy with the continuous does not give always
immediately the best possible direction to follow in the discrete, as
sometimes the peculiarities of such scenario should be explicitly taken into
account. The continuous is a limiting, and thus particular, case of the
discrete \cite{ruzzi1,barker,ruga2}, and not necessarily all the
conveniences of the continuous formalism are to be found in the discrete
approach. However, such conveniences (that is, established theoretical tools
with a commonly accepted physical interpretation) are often desirable, and a
lot of effort is usually made to achieve them in the discrete finite
dimensional scenario (a good example is the discussion about discrete Wigner
functions). As we argued above, in the search for a finite dimensional
squeezing operator, the position-momentum representation of the squeezing
operator is misleading, and another starting point had to be found.

As we have seen, such starting point is a complete set of states which is
parametrized by its width. The idea can be summarized in the following
lines: We construct an operator which maps unit width discrete oscillator
states on to discrete oscillator states of smaller (higher) width. As
discrete oscillator states are complete, we may conjecture that such
operator maps \textit{any} given state on to another of smaller (higher)
width, thus characterizing squeezing (stretching).

The first attempts to produce such operator, however, came up with something
which is not unitary, as the discrete oscillator states lack orthogonality.
We have seen, however, that if one works only with the lower discrete
oscillator states, unitarity can be reached in approximate fashion (in other
words, the squeezing transformation is approximately unitary in the
sub-space of the lower discrete oscillator states). This means that, as
states $|0;1\rangle $ has been successfully used to define a reference state 
\cite{maruga2} for finite dimensional coherent states, due exactly to its
convenient properties under DFT \cite{gama,rumaga1,maruga}, the squeezing
transformation over such reference state can be safely assumed to be unitary.

In the above cited reference, in fact, the properties of \textit{finite
coherent squeezed states}, obtained as finite displacements of a squeezed
reference state $|0;\xi \rangle $ are studied, and the first applications
of the here presented approach in quantum information theory (which may well
be the natural ground for applications of this formalism) are shown. Other
interesting field in which this formalism could be immediately implemented
is spin squeezing \cite{spinsqz}.

One could imagine that an alternative way of circumventing the non-unitarity
of the discrete oscillator states would be to obtain a new, orthogonal basis
by means of some orthogonalization process, like the Gram-Schmitd
procedure. Three such attempts have been extensively tested and none of them
has proved to be satisfactory, all for the same reason. The new, orthogonal,
sets of states, obtained trough all the attempted orthogonalization
processes, no longer have their dispersion as a monotonically increasing
function of the original squeezing parameter, so such a formalism, although
mathematically consistent, looses it physical significance. One of the
orthogonalization processes attempted was the Gram-Schmitd itself, the
second was a variation of it, and the third one was a method which is based
on the diagonalization of the overlap matrix.

One must stress that key feature of squeezing, aside from its scaling, shape
preserving effect on the wave function, is rather the consequence of this
scaling, which is to diminish (or enhance) the mean square deviation of a
given variable (in the case of a zero mean distribution). In fact, if taken
together with a displacement operator, the squeezing operator may squeeze
any distribution around its mean. The formalism here presented is focused
precisely at this point (our squeezing operator could be rather called root
mean square deviation diminishing operator, but, as Bohr once stated,
precision is the complementary counterpart of clarity), and it ultimately
gives one a procedure with which one may reduce the dispersion associated to
any quantum variable defined on a discrete domain.

As it is well known from quantum information theory, absolute fidelity of a
quantum teleportation protocol, if one does not want to destroy the original
system, can only be achieved if the original state is a member of an
orthogonal basis set. For a general state, it is clear that, the higher the
overlap of this state with a given basis element, the higher the fidelity of
the process. The squeezing transformation is completely capable to enhance,
or diminish, such fidelity. Therefore, one could argue that the ultimate end
of the here proposed finite dimensional squeezing operator may be to squeeze
out quantum information to the last q-bit.

\vspace*{1em}

\textbf{Acknowledgments:} The author thanks K. Zan and J. Belther for much
needed help with numerical calculations, and M.A. Marchiolli and D. Galetti
for lengthy and fruitful discussions.

\bigskip

\appendix\bigskip 

\section{Approximate unitarity}

In this appendix we may define the conditions in which we
can consider an operator of the following form 
\begin{equation}
\Xi _{u}(\xi )=\sum_{n=0}^{N_{l}-1}\ |n;\xi \rangle
(n;1|+\sum_{n=N_{l}}^{N-1}\ |n;1)\langle n;1|,
\end{equation}
as an ``approximately'' unitary operator. It's hermitian conjugate is directly calculated 
\begin{equation*}
\Xi _{u}^{\dagger }(\xi )=\sum_{n=0}^{N_{l}-1}\ |n;1)\langle n;\xi
|+\sum_{n=N_{l}}^{N-1}\ |n;1\rangle (n;1|,
\end{equation*}
so that, 
\begin{eqnarray*}
\Xi _{u}(\xi )\Xi _{u}^{\dagger }(\xi )
&=&\sum_{n=0}^{N_{l}-1}\sum_{m=0}^{N_{l}-1}\ |n;\xi \rangle (n;1|m;1)\langle
m;\xi |+\sum_{n=0}^{N_{l}-1}\sum_{m=N_{l}}^{N-1}\ |n;\xi \rangle
(n;1|m;1\rangle (m;1| \\
&&\sum_{n=N_{l}}^{N-1}\sum_{m=0}^{N_{l}-1}\ |n;1)\langle n;1|m;1)\langle
m;\xi |+\sum_{n=N_{l}}^{N-1}\ \sum_{m=N_{l}}^{N-1}|n;1)\langle
n;1|m;1\rangle (m;1|
\end{eqnarray*}
In order to study the situations in which the above operator is equal to the identity operator, the first needed criterion is 
 $\mathcal{V}_{l}^{\xi }$ and $%
\mathcal{V}_{h}^{\xi }$ are ``approximately'' orthogonal. We establish this
condition as

\begin{equation*}
|(m;\xi |n;\xi \rangle |^{2}<10^{-4},\qquad \mbox{for }n<N_{l}\leq m.
\end{equation*}
We now have 
\begin{equation*}
\Xi _{u}(\xi )\Xi _{u}^{\dagger }(\xi
)=\sum_{n=0}^{N_{l}-1}\sum_{m=0}^{N_{l}-1}\ |n;\xi \rangle (n;1|m;1)\langle
m;\xi |+\sum_{n=N_{l}}^{N-1}\ \sum_{m=N_{l}}^{N-1}|n;1)\langle
n;1|m;1\rangle (m;1|.
\end{equation*}
The second term is a different representation the identity operator of $%
\mathcal{V}_{h}^{\xi }\mathcal{\ }$\cite{arta}, whereas the first one is an
``approximate'' representation of the identity operator of $\mathcal{V}%
_{l}^{\xi }$ \ as long as $(n;1|m;1)\sim (n;\xi |m;\xi ).$ To establish
objectively this criterion we define that when 
\begin{equation*}
|(n;1|m;1)|^{2}-|(n;\xi |m;\xi )|^{2}<10^{-4}
\end{equation*}
the overlap $(n;\xi |m;\xi )$ is assumed not to depend on $\xi ,$ so under
this approximations

\begin{equation*}
\Xi _{u}(\xi )\Xi _{u}^{\dagger }(\xi )=\mathbf{\hat{1}}_{l}+\mathbf{\hat{1}}%
_{h}=\mathbf{\hat{1}}.
\end{equation*}
The conditions emerging from the relation $\Xi _{u}^{\dagger }(\xi )\Xi
_{u}(\xi )=\mathbf{\hat{1},}$ combined with the ones above can finally be
summarized in the conditions 
\begin{equation*}
|\langle m;\xi |n;\xi \rangle |^{2}<10^{-4},\qquad \mbox{for }n<N_{l}\leq m,
\end{equation*}
\begin{equation*}
|\langle n;1|m;1\rangle |^{2}-|\langle n;\xi |m;\xi \rangle |^{2}<10^{-4},
\end{equation*}
as the relations involving the dual vectors are consequences of those.


\begin{thebibliography}{99}

\bibitem{squeeztb}  H.M. Nussenzveig, Introduction to quantum optics, Gordon
and Breach Science Publishers, New York, 1973; M.O. Scully, M.S. Zubairy,
Quantum optics, Cambridge University Press, New York, 1997; M. Orszag,
Quantum optics, Springer, Berlin, 2000; W.P. Schleich, Quantum optics in
phase space, Wiley-VCH, Berlin, 2001; D.F. Walls, G.J. Milburn, Quantum
Optics (Springer Study Edition) Springer, Berlin, 1995; Quantum Optics, W.
Vogel , D-G. Welsch, Wiley-VCH, Berlin, 2006.

\bibitem{squeezra}  D. Walls, Nature 306, 141, 1983; V. V. Dodonov,
`Nonclassical' states in quantum optics: a `squeezed' review of the first 75
years, J. Opt. B: Quantum Semiclass. Opt. \textbf{4} (2002) R1-R33; S.L.
Braunstein, P. van Loock, Quantum information with continuous variables,
Rev. Mod. Phys. \textbf{77} (2005) 513.

\bibitem{noc}  D. Dieks, Phys. Lett. \textbf{92A}, (1982) 271; W.K. Wootters
and W. H. Zurek, Nature \textbf{299}, (1982) 802.

\bibitem{qsr} E. Bagan, M. Baig and R. Mun\~oz-Tapia, Phys. Rev. Lett \textbf{89} (2002) 277904.



\bibitem{vourdas}  A. Vourdas, Quantum systems with finite Hilbert space,
Rep. Prog. Phys. \textbf{67} (2004) 267.

\bibitem{numb}  G. E. Andrews, Number Theory, Dover, New York, 1994.



\bibitem{schw}  J. Schwinger, Proc Natl Acad Sci, 46 : 570 1960 .

\bibitem{ruzzi1}  M. Ruzzi, Schwinger, Pegg and Barnett approaches and a
relationship between angular and Cartesian quantum descriptions, J. Phys. A:
Math. Gen. \textbf{35} (2002) 176.


\bibitem{eujmp}  M. Ruzzi, Jacobi $\vartheta $-functions and discrete
Fourier transforms, J. Math. Phys. \textbf{47} (2006) 063507.

\bibitem{vilenkin}  N.J. Vilenkin, A.U. Klimyk, Representation of Lie groups
and special functions: Simplest Lie groups, special functions and integral
transforms, Kluwer Academic, Dordrecht, 1992.


\bibitem{mehta}  M.L. Mehta, Eigenvalues and eigenvectors of the finite
Fourier transform, J. Math. Phys. \textbf{28} (1987) 781.


\bibitem{natig}  M. N. Atakishiyev, A. U. Klimyk, The factorization of a q-difference equation for continuous q-Hermite polynomials,
J. Phys. A: Math. Theor. \textbf{40}  (2007) 9311.

\bibitem{maruga2}  M.A. Marchiolli, M. Ruzzi, D. Galetti, Discrete squeezed
states for finite-dimensional spaces, Phys. Rev. A \textbf{76} (2007) 032102.


\bibitem{bellman}  R. Bellman, A brief introduction to theta functions, Holt
Rinehart and Winston, New York, 1961.

\bibitem{arta}  E. Artacho, L.M. del Bosch, Nonorthogonal basis sets in
quantum mechanics: Representations and second quantizatio, Phys. Rev. A,
textbf{43} (1991) 5770.

\bibitem{weyl}  H. Weyl, The Theory of Groups and Quantum Mechanics, Dover,
New York, 1950.

\bibitem{barker}  L Barker, Continuum quantum systems as limits of discrete
quantum systems, I: State vectors J. Func. Anal. \textbf{186} (2001) 153;
Continuum quantum systems as limits of discrete quantum systems: II. State
functions, J. Phys. A Math. Gen \textbf{34} (2001) 4673; Continuum quantum
systems as limits of discrete quantum systems. III. Operators, J. Math.
Phys. \textbf{42} (2001) 4653; Continuum quantum systems as limits of
discrete quantum systems. IV. Affine canonical transforms, J. Math. Phys 
\textbf{44} (2003) 1535.

\bibitem{ruga2}  M Ruzzi, D. Galetti, Schwinger and Pegg-Barnett
approaches and a relationship between angular and Cartesian quantum
descriptions: II. Phase spaces J. Phys A: Math. Gen \textbf{35} (2002) 4633.


\bibitem{gama}  D. Galetti, M.A. Marchiolli, Discrete coherent states and
probability distributions in finite-dimensional spaces, Ann. Phys. \textbf{%
249} (1996) 454.

\bibitem{rumaga1}  M. Ruzzi, M.A. Marchiolli, D. Galetti, Extended
Cahill-Glauber formalism for finite-dimensional spaces: I. Fundamentals, J.
Phys. A: Math. Gen. \textbf{38} (2005) 6239;

\bibitem{maruga}  M.A. Marchiolli, M. Ruzzi, D. Galetti, Extended
Cahill-Glauber formalism for finite-dimensional spaces: II. Applications in
quantum tomography and quantum teleportation, Phys. Rev. A \textbf{72}
(2005) 042308.


\bibitem{spinsqz}  M. Kitagawa and M. Ueda, Phys. Rev. A 47, 5138 (1993). %
\end{thebibliography}
\end{document}